\documentclass{aa}

\usepackage{graphicx}
\usepackage{txfonts}
\usepackage{hyperref}
\usepackage[dvipsnames]{xcolor}

\newcommand{\kpc}{~\mathrm{kpc}}
\newcommand{\Mpc}{~\mathrm{Mpc}}
\newcommand{\Mo}{~\mathrm{M}_\odot}
\newcommand{\dex}{~\mathrm{dex}}

\newcommand{\Mopcmt}{~\mathrm{M}_\odot\,\mathrm{pc}^{-2}}

\newcommand{\kpcmt}{~\mathrm{kpc}^{-2}}
\newcommand{\magamt}{~\mathrm{mag}\,\mathrm{arcsec}^{-2}}

\begin{document} 

   \title{An accurate measure of the size of dark matter haloes using the size of galaxies}

   \subtitle{}

   \author{Claudio Dalla Vecchia
          \inst{1,2}\fnmsep\thanks{email: claudio.dalla.vecchia@iac.es}
          \and
          Ignacio Trujillo
          \inst{1,2}
          }

   \authorrunning{C.~Dalla Vecchia and I.~Trujillo}
   
   \institute{Instituto de Astrof\'isica de Canarias, C/ V\'ia L\'actea s/n, 38205 La Laguna, Spain
         \and
             Departamento de Astrof\'isica, Universidad de La Laguna, Av.~del Astrof\'isico Francisco S\'anchez s/n, 38206 La Laguna, Spain
             }

   \date{Received 28 June 2025 / Accepted 17 October 2025}

     \abstract{The physically motivated definition of galaxy size proposed recently, linked to the farther location of the in situ star formation, considerably reduces the scatter of the galaxy mass--size relation and provides a viable method to infer the galaxy stellar mass from its size. We provide a similar relation correlating the size of galaxies with the size of their dark matter haloes by leveraging the small scatter of the aforementioned relation. We analysed the simulated galaxies of the two main cosmological volumes of the EAGLE simulations and computed the size of the galaxies and their mass
     when mimicking the observational analysis. For central galaxies, we computed the relation between galaxy size and halo size. We show that the simulated galaxies reproduce the observed stellar mass--size relation's normalisation and slope. The scatter of this relation, $0.06\dex$, matches the intrinsic scatter measured in observation. We then computed the correlation between galaxy size and halo size and found that the relation is steeper than when using the half-mass radius as a measure of size, with the scatter $(0.1\dex)$ being a factor of two smaller than the observed relation. As well, the galaxy-to-halo mass relation derived from the simulations provides a factor of two better scatter than
     the observed scatter. This opens the possibility of measuring the size of dark matter haloes with greater accuracy (less than 50\%, i.e.~around  six times better than using the effective radius) by using only deep imaging data.}


   \maketitle

\section{Introduction}
\label{sec:intro}

A physically motivated definition of galaxy size, given by the farthest radial location where gas could efficiently collapse and turn into stars, was proposed by \citet{Trujillo2020}. As a proxy for measuring this position, these authors suggest using the radial location of a stellar mass density contour at $1\Mopcmt$, denoted $R_1$.  This measure is consistent with the gas density threshold for star formation derived theoretically by \citet{Schaye2004}. The use of $1\Mopcmt$ was calibrated using deep observations of Milky Way-sized galaxies
at $z\sim0$,
and it provides a more representative boundary for galaxies than traditional metrics such as the effective radius, $R_\mathrm{e}$ \citep{Sersic1968}. Using this definition over a wide range of stellar masses (from $10^7$ to $10^{12}~\text{M}_\odot$) significantly reduces the scatter in the stellar mass--size relation to about 0.06~dex, thus outperforming conventional measures such as $R_\mathrm{e}$ and the Holmberg radius, $R_\mathrm{H}$ \citep{Holmberg1958}. 

Even though $R_1$ is only a proxy to characterise the physical radius where in situ star formation is or has been taking place, the enormous reduction in the scatter of the stellar mass--size relation underscores the robustness of $R_1$ as a galaxy size metric. The results of using $R_1$ also reveal clear trends within the stellar mass--size relation. Galaxies with stellar masses between $10^7$ and $10^{11}~\text{M}_\odot$ follow a consistent power-law relationship with a slope of around 1/3. Across the range of galaxy stellar masses studied, the scatter is consistently small ($\sim$0.06 dex), providing an accurate measure of the galaxy's stellar mass by knowing only its physical size.

Observed disc galaxies follow the relation $R_1\propto M_\star^{1/3}$ over several orders of magnitude in stellar mass. A similar relation has been proposed for neutral hydrogen by \citet{Broeils1997}. The mass--size relation for neutral hydrogen is $M_{\text{H\,\textsc{i}}}\propto D_{\text{H}\,\textsc{i}}^{2}$, where $D_{\text{H}\,\textsc{i}}$ is the diameter of the neutral hydrogen disc of a spiral galaxy. \citet{Wang2016}, using observations of 500 galaxies, showed that this correlation holds irrespective of galaxy luminosity, fraction of neutral gas, and morphology, with a scatter of $0.06\dex$. This points to the non-trivial interplay of gas accretion, star formation efficiency, and stellar feedback in maintaining the two mass--size relations over several orders of magnitude in stellar mass.

\citet{Chamba2022} tested the validity of the hypothesis of using $R_1$ as an indicator of the radius up to which in situ star formation has occurred. To do this, they identified the position of the stellar edge of galaxies, $R_\mathrm{edge}$,
as the outermost cut-off in their stellar mass radial profile using deep multi-band optical imaging; verified that it corresponds to a change in the radial colour profile and surface brightness profile;
and measured the stellar mass surface density it corresponds to.
\citet{Chamba2022} find that $R_\mathrm{edge}\sim R_1$ for galaxies with masses similar to the Milky Way \citep[see also][]{Golini2025}.
For lower-mass galaxies (dwarfs) or more massive galaxies (ellipticals), the surface mass density of the galaxy edge is slightly different from $1\Mopcmt$. It is smaller by a factor of about two for low-mass galaxies and larger by a factor of about three for massive elliptical galaxies. Be that as it may, within a range of five orders of magnitude in stellar mass, the approximation of using $R_1$ as an indicator of galaxy size is quite good, and its practical implementation is straightforward. One of the main results of \citet{Trujillo2020} is that as long as the definition of galaxy size includes most of the stellar mass, the mass--size relation should have a small scatter \citep[e.g.][]{Miller2019}. \citet{SanchezAlmeida2020} has shown that for a fixed stellar mass, stellar profiles with different S\'ersic indexes all cross at a similar stellar mass surface density, and they concluded that this explains the small scatter when using $R_1$ as the size.

In a subsequent paper, \citet{Buitrago2024} studied the evolution of the stellar mass--size relation with redshift for a sample of Milky Way-like galaxies, directly measuring $R_\mathrm{edge}$ (instead of $R_1$) of galaxies up to $z=1$. They showed that the edge is located at higher stellar surface densities at a higher redshift and that disc galaxies of the same stellar mass are smaller with increasing redshift, with their size inversely proportional to $(1+z)$. This shows that using $R_1$ as a proxy for galaxy size is a valid option only at $z\sim0$. The dependence of galaxy size on redshift reveals the connection between the growth of the dark matter halo and the galaxy disc, as the former grows in size with the same redshift scaling \citep{Navarro1997}.

In a companion paper, \citet{Arjona2025} show that when using $R_1$ as the definition of size, galaxies extracted from a variety of zoom-in simulations performed with different models for galaxy formation all share the same mass--size relation. When studying its evolution in simulations, the relation does not change with redshift; rather, the galaxies move on the relations as they grow in mass and size.
The difference between this and the observations of \citet{Buitrago2024} is that the stellar surface density threshold is kept fixed with redshift in the analysis, whilst observations show that the stellar surface density at which disc truncation is observed is larger at a higher redshift. What the simulations have in common is that their star formation models all try to reproduce the observed (local) Kennicutt-Schmidt law and its surface density threshold for star formation \citep{Kennicutt1998,Martin2001,Bigiel2008}, which is assumed to hold at any redshift. Combining this with a fixed stellar surface density may result in galaxies of the same mass having the same $R_1$ at any redshift.

Given the close connection between the new definition of galaxy size and the physical process of in situ star formation, as well as the observation that the size thus defined scales with the expected evolution of dark matter halo size, in this paper we explore the correlation between galaxy size, $R_1$, and dark matter halo size, $R_{200}$, defined as the radius of the sphere with an average mass density 200 times the critical density of the Universe. We find that the simulations reproduce the observed relation between stellar mass and size remarkably well. Moreover, they predict that the relation between $R_1$ and $R_{200}$ is extremely tight so that when given $R_1$, one can estimate $R_{200}$ with an uncertainty of less than 50\%. This is a factor of more than two better than the predicted relation between the effective radius and $R_{200}$ \citep{Kravtsov2013} and the predicted stellar-to-halo mass relation for observed galaxies when the galaxy size is given by the radius containing 80\% of the light \citep{Mowla2019}.

In this work, we analyse the two main cosmological hydrodynamical simulations of the EAGLE project \citep{Schaye2015,Crain2015} for the stellar mass range $\approx 10^{7.5}$--$10^{12}\Mo$. We describe the data used in the simulations and the analysis in the following section. In Section \ref{sec:results}, we compare the results from the simulations with observations. In Section \ref{sec:halosize}, we link the size of the galaxies to the size of their halo, showing that this correlation can predict the halo mass and size with a higher accuracy when compared to using the half-mass radius or $R_\mathrm{e}$. A discussion and our conclusions are presented in Section \ref{sec:conclusions}.

\section{Analysis of the simulations}
\label{sec:analysis}

\subsection{Simulations and simulation data}
\label{subsec:simdata}

The simulations employed in this work were performed with the EAGLE model for galaxy formation and evolution. For more information on the EAGLE model and its calibration against global relations of the observed galaxy population (galaxy stellar mass function, stellar mass--size relation, and stellar mass-black hole mass relation), we refer to \citet{Schaye2015} and \citet{Crain2015}. For more details on the numerical algorithms describing the photo-ionisation equilibrium cooling, star formation, stellar evolution, stellar feedback, black hole growth, and feedback we refer to \citet{Schaye2008}, \citet{Wiersma2009a,Wiersma2009b}, \citet{DallaVecchia2012}, \citet{Rosas-Guevara2015}, respectively. A description of the benefits of the hydrodynamic scheme used in the EAGLE model can be found in \citet{Schaller2015b} \citep[see also][]{Durier2012}. For the analysis presented here, the raw particle data at $z=0$ and the public database\footnote{\url{http://icc.dur.ac.uk/Eagle/}} were used \citep{McAlpine2016, EAGLE17}. 

The EAGLE model implements the sub-grid prescription for star formation of \citet{Schaye2008}. Briefly, the model is based on the conversion of the Kennicutt-Schmidt empirical law \citep{Kennicutt1998} that correlates the surface rate of star formation
to the surface density of gas, to a volumetric law that expresses the rate of star formation as a function of gas pressure. The density threshold above which the gas is able to form stars is assumed to vary with the metallicity of the gas \citep{Schaye2004}.
Therefore, star formation in EAGLE occurs from the threshold density set by the metallicity of the gas and beyond. This threshold decreases with increasing metallicity. Details on the numerical implementation of the dependence on metallicity of the density threshold for star formation can be found in \citet{Schaye2015}. We emphasise that the theoretical predictions of \citet{Schaye2004} are used as a starting hypothesis for the definition of galaxy size by \citet{Trujillo2020}.

\begin{figure*}
    \centering
    \includegraphics[width=\textwidth]{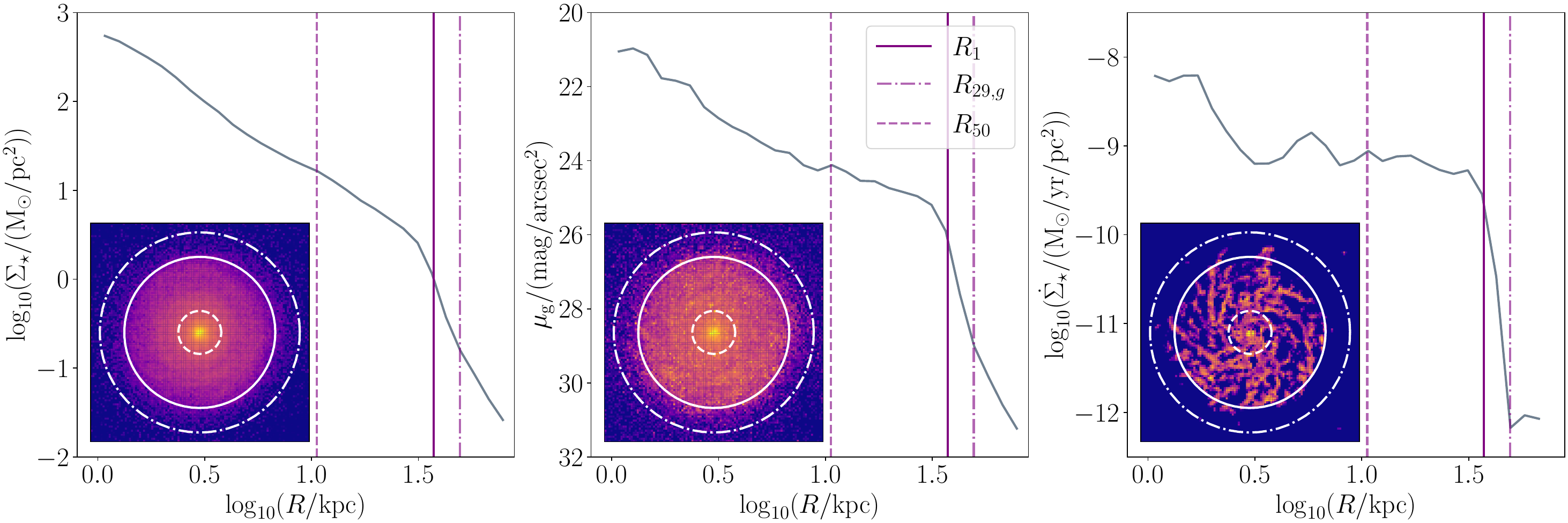}
    \caption{From left to right, projected density maps and radial profiles of stellar mass, stellar luminosity, and gas star formation rate for a simulated spiral galaxy of stellar mass $3.8\times 10^{10}\Mo$ selected from the recalibrated simulation. The vertical dashed, solid, and dot-dashed lines mark the positions of $R_{50}$, $R_1$, and $R_{29,g}$, respectively. For this star forming galaxy, $R_1$ is a better measure of the size of the galaxy for its proximity to the edge (truncation radius) of the disc. The edge of the disc coincides with a net drop in the star formation rate surface density, the starting hypothesis for the new definition of galaxy size.}
    \label{fig:profiles}
\end{figure*}

The post-processing of the simulations was performed with the friends-of-friends algorithm \citep[e.g.][]{Einasto1984,Davis1985} and the SubFind algorithms \citep{Springel2001,Dolag2009}, which define the distribution of overdense regions in the simulated cosmological volumes and their substructure. We further post-processed the $z=0$ outputs of the EAGLE simulations to compute dust-attenuated stellar absorption spectra that we convolved with filter response curves and integrated in the wavelength bands of the Sloan Digital Sky Survey \citep{Abazajian2003}. We employed the EMILES stellar spectra library \citep{Vazdekis2016} and the same methodology described in \citet{Negri2022}. More details are given in that work.

We analysed the $z=0$ output of two EAGLE simulations. The reference EAGLE volume of $(100\Mpc)^3$ -- the `reference' simulation -- initially contained $2\times 1504^3$ dark matter and gas particles, had a physical spatial resolution of $\epsilon=0.7\kpc$, had particle masses of $m_\mathrm{gas}=1.81\times 10^6\Mo$ and $m_\mathrm{dm}=9.70\times 10^6\Mo$, for gas and dark matter, respectively. The recalibrated EAGLE volume of $(25\Mpc)^3$ -- the `recalibrated' simulation -- initially contained $2\times 752^3$ dark matter and gas particles, had a physical spatial resolution of $\epsilon=0.35\kpc$, and had particle masses of $m_\mathrm{gas}=2.26\times 10^{5}\Mo$ and $m_\mathrm{dm}=1.21\times 10^6\Mo$.
For the recalibrated simulation, the EAGLE galaxy formation model was calibrated in order to match the same global relations at $z=0$ reproduced within the largest volume \citep{Crain2015}. The recalibrated simulation was used to extend the analysis down to galaxy stellar masses $\approx 2\times 10^7\Mo$ (about $1\dex$ smaller stellar masses than in the reference simulation), after restricting the minimum stellar mass to about 100 stellar particles. Finally, the initial conditions for the aforementioned simulations were generated with the cosmological parameters inferred by the \citet{PlanckXVI}: $\Omega_\mathrm{0}=0.307$, $\Omega_\Lambda=0.693$, $\Omega_\mathrm{b}=0.04825$, $h=0.6777$, $\sigma_8=0.8288$, and $n_\mathrm{s}=0.9611$.

\subsection{Determination of galaxy size and stellar mass}
\label{subsec:size}

We describe here how we measured the size of the simulated galaxies. The $R_1$ radius has been defined as the projected distance from the centre of the galaxy to the location where the stellar surface density reaches $1\Mopcmt$. We determined $R_1$ by first rotating the galaxy face-on using the angular momentum vector of the stellar component, thus avoiding the uncertainties introduced by de-projecting the stellar distribution. Because of the relatively low spatial and mass resolution of the simulations, we did not calculate maps of the stellar mass surface density as usually done in observational work. Indeed, given the stellar particle mass of about $10^6\Mo$ for the reference simulation, at the threshold surface density of $1\Mopcmt$, the particle number density would be about $1\kpcmt$, making any projected surface density very noisy. We therefore opted to use the stellar surface density profile derived by averaging the projected stellar mass in circular annuli. We set the centre of the galaxy at the position of the minimum of the gravitational potential of the halo, as provided by the sub-halo finder. This is a good estimate of the centre of the galaxy, as the offset of the stellar component with respect to the dark matter halo is less than the spatial resolution of the simulation for the majority of galaxies \citep{Schaller2015a}. $R_1$ was then derived by interpolating the surface density profile at the position of the threshold surface density.

We determined the galaxy stellar mass as the projected stellar mass within a radius of constant surface brightness. This was done in order to consistently include the same bias introduced by observers in determining the relation. Indeed, there is no direct correlation between the measurement of $R_1$ and $M_\star$ in \citet{Trujillo2020}.
$R_1$ was derived by converting the surface brightness into surface stellar mass density, while
$M_\star$ was derived by integrating the
stellar mass above within the surface brightness threshold of $29\magamt$ in the $g$ band. This corresponds to a stellar surface density threshold lower than $1\Mopcmt$. In order to calculate the position of the threshold in surface brightness, we converted the stellar mass surface density by inverting equation (1) of \citet{Bakos2008}:
\begin{equation}
    \log_{10} \Sigma_\star= \log_{10} (M/L)_\lambda - 0.4(\mu_\lambda - m_{\mathrm{abs},\sun,\lambda}) + 8.629,
\end{equation}
where $\Sigma_\star$ is the stellar surface density in$\Mopcmt$, $M/L$ is the mass-to-light ratio at the wavelength $\lambda$, and $m_{\mathrm{abs},\sun,\lambda}$ is the absolute magnitude of the Sun at the same wavelength. We calculated $M/L$ with the Sloan Digital Sky Survey $g$ band magnitude that we computed in the post-processing of the simulations for each stellar particle. From here on, we denote the projected stellar mass within $R_{29,\mathrm{g}}$ as $M_\star$.

For each galaxy, we also calculated the stellar half-mass radius, $R_{50}$, as the radius containing half the projected stellar mass. This radius is a good proxy for $R_\mathrm{e}$. The computation of $R_1$, $R_{50}$, and $R_{29,\mathrm{g}}$ was performed for all simulated galaxies with a mass larger than 100 initial gas particle masses ($2.26\times 10^7$ and $1.81\times 10^8\Mo$ for the recalibrated and reference simulations, respectively). However, the algorithm failed to converge for a fraction of galaxies that are barely resolved or have a disturbed morphology. We marked them and did not use them in the analysis. The final selection of the samples of galaxies used in the analysis is described in Appendix~\ref{app:sample}.

We show in Figure~\ref{fig:profiles} one example of a simulated spiral galaxy with a stellar mass of $3.8\times 10^{10}\Mo$.
From left to right, we show the radial profiles for the projected stellar surface density, the surface brightness, and the star formation surface density.
The insets are the corresponding maps and have an arbitrary logarithmic colour scale.
The circles in the maps and the vertical lines in the profiles plots indicate the values of the half-mass radius, $R_{50}$ (dashed line); $R_{29,\mathrm{g}}$ (dash-dotted line); and the galaxy size, $R_1$ (solid line).
For the galaxy shown in Figure~\ref{fig:profiles}, $R_1$ is an excellent proxy for measuring the edge of the star forming disc and thus a good approximation of the physically motivated definition of galaxy size.  
The stellar disc truncation is close to $R_1$, and the gas surface density steeply declines beyond $R_1$ and so does the star formation surface density.
If the galaxy ceased to form stars for the rest of its evolution and without contribution to its stellar mass from mergers, the edge of the in situ star formation would still be marked by $R_1$.

\section{Results}
\label{sec:results}

\begin{figure}
    \centering
    \includegraphics[width=0.48\textwidth]{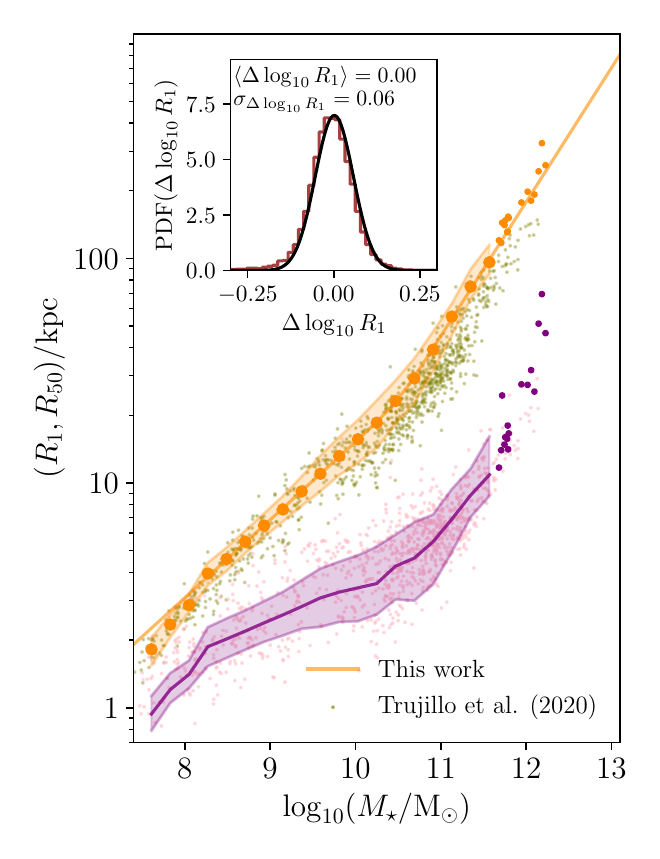}
    \caption{Correlation between $R_1$ and stellar mass for all galaxies in the sample (orange symbols). The median values in each bin are plotted together with the 16\% and 84\% quantiles (shaded area). For bins with fewer than ten galaxies, single galaxies are plotted as small dots. The orange solid line depicts the double power-law fit for $M_\star>10^{8.6}\Mo$ (extrapolated below that mass).
    We show the relation between $R_{50}$ and stellar mass for all galaxies in the sample in purple.
    For visual comparison, the galaxies in the sample of \protect\citet{Trujillo2020} are represented with small dots, both $R_1$ and $R_{50}$.
    The distribution of the residuals around the best-fit linear relation for all galaxies is shown in the inset. The distribution has been fitted with a Gaussian (solid line) with a dispersion of $\sigma_{\Delta\log_{10}R_1}=0.06~\mathrm{dex}$.}
    \label{fig:msra}
\end{figure}

The results of the study of the galaxy mass--size relation in simulations is presented here. We show in figure~\ref{fig:msra} the correlation between $R_1$ and galaxy stellar mass for all galaxies in the sample, central and satellite.\footnote{\citet{Trujillo2020} did not discriminate between central and satellite galaxies.}
The sample was selected as described in Appendix~\ref{app:sample}.
We emphasise that even when using the same definition of $R_1$, the analysis of the simulated data differs substantially from that of observational data.
The large dots in the figure are the median values of $R_1$ measured as described in Section~\ref{subsec:size}, while the small dots are individual galaxies where the stellar mass bin contains fewer than ten galaxies.
The shaded area represents the 16\% and 84\% quantiles in each bin.
The simulated relation follows the trend of observations, with large slopes for stellar masses below $10^{8.5}\Mo$ and above $10^{10.5}\Mo$ with respect to the the mass range in between.
As mentioned in the introduction, \citet{Trujillo2020} explains that this is due to the fixed surface density threshold employed in their analysis.
The same trend is shown in the figure of Appendix~\ref{app:censat}, where we show the calculated relation for the samples of central\footnote{A central galaxy is defined by \textsc{subfind} as the galaxy sitting at the potential minimum of its friends-of-friends halo.} and satellite galaxies.

For comparison, in Figure~\ref{fig:msra} we also plot the median distribution of projected stellar half-mass radii, $R_{50}$, with its 16\% and 84\% quantiles (purple solid line and shaded area). The scatter around the median is more than double of that for $R_1$. Moreover, the correlation between $R_{50}$ and stellar mass is very weak in the mass range $10^{8.5}<M_\star/\Mo<10^{10.5}$, where the mass of galaxies with similar $R_{50}\approx 2-3\kpc$ spans over two orders of magnitude. The weak correlation and its large scatter provide a relation with little information on the properties of galaxies. However, for a given stellar mass, the scatter around $R_{50}$ can provide insight into how gas has been redistributed by feedback processes within a star forming galaxy \citep[e.g.][]{Crain2015,Rohr2022}.  

In order to provide an analytic fit to the relation, we chose the extended double power-law function defined as follows:
\begin{equation}
    \label{eqn:powerlaw}
    y=A\left(\frac{x}{x_\mathrm{s}}\right)^\alpha\left(1 + \left(\frac{x}{x_\mathrm{s}}\right)^s \right)^{(\beta - \alpha)/s},
\end{equation}
with free parameters $(A,x_\mathrm{s},\alpha,\beta,s)$. Parameter $A$ sets the normalisation, $x_\mathrm{s}$ is the scale at which the slope transitions between the asymptotic values $\alpha$ and $\beta$, and $s$ is the strength of this transition. Here, we assumed $(x,y,x_\mathrm{s})\equiv(M_\star, R_1, M_\mathrm{s})$.
We then fit the median values for $M_\star>10^{8.5}\Mo$, excluding the low-mass end drop of $R_1$ with decreasing mass.\footnote{We refer the the reader interested in the low-mass regime to \citet{Arjona2025} for a similar study employing high-resolution simulations.}
The $\chi$-square minimisation gives the following set of parameters for the best-fit function: $(A,M_\mathrm{s},\alpha,\beta,s)\approx(22.8\kpc, 3.21\times10^{10}\Mo,0.35,0.60,4.25)$. The asymptotic slopes, $\alpha=0.35$ and $\beta=0.60$, are remarkably close to the observed values given by \citet{Trujillo2020} (and the high-mass slope given by \citet{Mowla2019}).

In Appendix~\ref{app:censat}, we provide similar fits for central and satellite galaxies and a table with the double power-law fit parameters and their errors for the three sub-samples of galaxies.
As a check of the quality of the fit to the binned values, we also calculated the best-fit linear function to individual galaxies within the mass range $10^{8.5}\leq M_\star/\mathrm{M}_\odot\leq 10^{10.5}$ (dashed line), which yielded the correlation
\begin{equation}
    \log_{10}\left(\frac{R_1}{\mathrm{kpc}}\right) = (0.343\pm 0.001)\log_{10}\left(\frac{M_\star}{\mathrm{M}_\odot}\right) - (2.258\pm 0.009)\,,
    \label{eqn:msrc}
\end{equation}
which successfully matches the asymptotic slope of the fit to the median.

The slope of the size-mass relation increases to a larger value above the mass $M_\mathrm{s}\approx 10^{10.5}\Mo$. In numerical modelling, this is a characteristic mass that marks the transition between supernova and black hole feedback dominance for several numerical models \citep[e.g][]{Bower2017,McAlpine2017,Pillepich2018}. Observations show a similar characteristic mass. \citet{Trujillo2020} give a value close to $10^{10.8}\Mo$, whilst \citet{Mowla2019} quote a mass of $10^{10.2}\Mo$. In both cases, the characteristic stellar mass marks the morphological transition from disc to elliptical galaxies, which is, from theoretical studies, closely connected to the transition between stellar and black hole feedback and to the transition in the process of stellar mass growth from gas accretion to dry mergers with increasing halo mass.

Finally, we show in the inset in Figure~\ref{fig:msra} the distribution of residuals around the best-fit stellar mass--size relation and in the same mass range used for the calculation of the best-fit analytic function for the full sample of galaxies. The distribution is well represented by a Gaussian with a central value of $\langle\Delta\log_{10}R_1\rangle = 0.00~\mathrm{dex}$ and a dispersion of $\sigma_{\Delta\log_{10}R_1}=0.06~\mathrm{dex}$, in agreement with the intrinsic scatter measured by \citet{Trujillo2020}. We show in the next section that this very small scatter yields a tighter correlation between galaxy and halo size than what was inferred by \citet{Kravtsov2013} and a smaller scatter in the stellar-to-halo mass relation than that observed by \citet{Mowla2019} with their definition of galaxy size.

\begin{figure}
    \centering
    \includegraphics[width=0.48\textwidth]{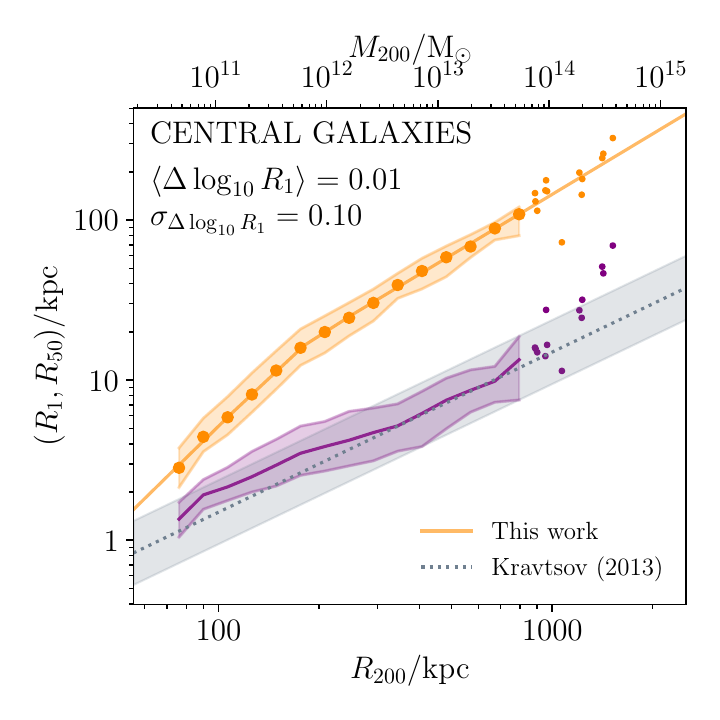}
    \caption{Correlation between $R_1$ and $R_{200}$ for central galaxies. Median (dots and solid lines) and 16\% and 84\% quantiles (shaded areas) are shown. We quote in the legend the average of the scatter for the entire sample of central galaxies and with respect to the analytic fit
    as well as the dispersion around the mean. The purple line and shaded area are the relation between $R_{50}$ and $R_{200}$ and its 16\% and 84\% quantiles. The dotted line is the
    linear correlation proposed by \citet{Kravtsov2013} for early-type galaxies, $R_{50}=0.015R_{200}$, 
    with a scatter of $0.2~\dex$ (shaded area).
    The top x-axis gives the corresponding halo mass, $M_{200}$.}
    \label{fig:r1r200}
\end{figure}

\section{Correlation between galaxy and halo sizes}
\label{sec:halosize}

\begin{figure}
    \centering
    \includegraphics[width=0.48\textwidth]{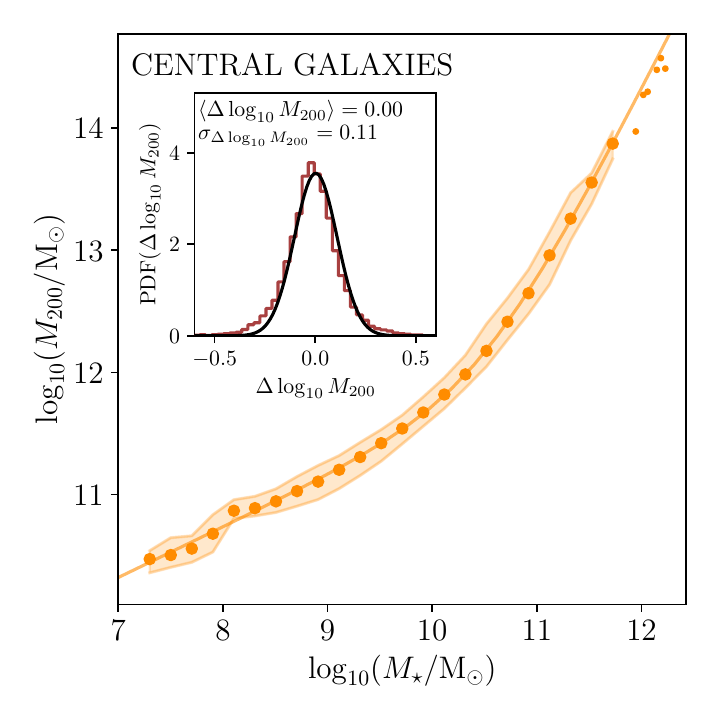}
    \caption{Correlation between $M_{200}$ and $M_\star$ for central galaxies. Median (dots and solid lines) and 16\% and 84\% quantiles (shaded areas) are shown. We quote in the legend the average of the scatter for the entire sample of central galaxies and with respect to the analytic fit, together with dispersion around the mean. We plot in the inset the distribution of residuals for the analytic fitting function and all the galaxies in the sample.}
    \label{fig:msm200}
\end{figure}

In this section, we present the main result of this work, which is the study of the correlation between galaxy and halo sizes. We show in Figure~\ref{fig:r1r200} the relation between $R_1$ and $R_{200}$.
The top horizontal axis gives the mass of the halo, $M_{200}$, proportional to $R_{200}^3$. By definition, $M_{200}$ and $R_{200}$ are the mass and radius of a sphere centred on the potential minimum and with an average density 200 times the critical density of the Universe: $M_{\delta}=(4/3)\pi\delta\rho_\mathrm{crit} R_\delta^3$, with $\delta=200$.

We restricted the sample to only central galaxies because the calculation of the mass and radius of their haloes is more reliable. The data distribution is shown as the median in equally spaced logarithmic bins (orange dots), with the orange shaded area delimiting the 16\% and 84\% quantiles. The correlation between $R_1$ and stellar mass can be described by a double power-law function. We employed the extended double power-law function of equation~\ref{eqn:powerlaw}, with $(x,y,x_\mathrm{s})\equiv(R_{200},R_1,R_\mathrm{s})$,
whose $\chi$-square minimisation fit provides the parameter values $(A,R_\mathrm{s},\alpha,\beta,s)\approx(16.4\kpc, 175.8\kpc,2.04,1.25,15.1)$. The scale radius, $R_\mathrm{s}$, corresponds to the halo mass of about $10^{11.8}\Mo$. All fitted parameters with errors are given in Table~\ref{app:tab2}.
Given the tight correlation between $R_1$ and $M_\star$, the above correlation can be interpreted
as the stellar-to-halo mass relation \citep[see][for more details, and later in this section for the comparison with observations]{Arjona2025}. 

For comparison, in Figure~\ref{fig:r1r200} we also plot the relation between $R_{50}$ and $R_{200}$ (purple line and shaded area) and the corresponding linear relation of \citet{Kravtsov2013} (dotted line). The $R_{50}=0.015 R_{200}$ relation was derived using a large sample of observed galaxies with stellar masses between $10^5$ and $10^{12}\Mo$, and it qualitatively agrees with the EAGLE simulation data. The scatter measured by \citet{Kravtsov2013}, $\approx 0.2\dex$, is twice that of the simulated galaxies, $\sigma_{\Delta \log_{10}R_1}=0.1\dex$.

\citet{Kravtsov2013} assumed that the relation between galaxy and halo sizes is linear, with a well-defined proportionality for all galaxies and their haloes \citep[see also][for a similar hypothesis]{Somerville2018}. The reasoning is that simulations show a very tight distribution of the spin parameter, $\lambda$,  of dark matter haloes at any scale \citep[e.g.][]{Peebles1969,Bullock2001}, and if the halo angular momentum dictates the size of the galaxy's disc, there should be a well-defined proportionality between the distribution of stars in the disc and the size of its dark matter halo. At a fixed stellar mass, variations of the stellar half-mass radius should reflect the dispersion around the average value of $\lambda$.
Although we found some indication that the scatter in the mass--size relation may correlate with $\lambda$ at a fixed stellar mass (not shown in this work), this correlation is weak.

The $R_1$--$R_{200}$ relation asymptotically approaches the slope $\beta\approx 1.25$ for halo masses larger than $\approx10^{11.5}\Mo$, which is steeper than what is inferred for observed galaxies. We conclude that given the small scatter and prominent slope in the $R_1$--$R_{200}$ relation, the physical size of observed galaxies can be used to infer the size (and mass) of their host haloes with noticeable accuracy. For a galaxy such as the Milky Way, using $R_1$ to infer $R_{200}$ would be six times more precise than using $R_{50}$.

As a final exercise, we analysed the stellar-to-halo mass relation, shown in Figure~\ref{fig:msm200}. We computed the best-fit parameters for the double power-law function of equation~\ref{eqn:powerlaw}, with $(x,y,x_\mathrm{s})\equiv(M_\star,M_{200},M_\mathrm{s})$, which gives the values $(A,M_\mathrm{s},\alpha,\beta,s)\approx(5.8\times10^{11}\Mo, 3.13\times10^{10}\Mo,0.41,1.69,0.98)$, confirming the change of slope at $M_\mathrm{s}\approx 10^{10.5}\Mo$. The residual distribution is shown in the inset of Figure~\ref{fig:msm200}, with the calculated dispersion around the mean of $0.11\dex$, which is half the value reported by \citet{Mowla2019} in their work.

\section{Discussion}
\label{sec:discussion}

The definition of galaxy size of \citet{Trujillo2020} is physically motivated by the theoretical prediction that the onset of star formation requires specific conditions for the gas to collapse into star forming clouds \citep{Schaye2004}. Accretion of gas into the galaxy and its subsequent cooling provide the material for the galaxy to grow both in stellar mass and size through star formation. Therefore, the physical edge of a star forming galaxy can be defined by the extension of its in situ star formation activity. The threshold gas surface density for star formation given by \citet{Schaye2004} can be converted into a stellar surface density that, even if the galaxy stops forming stars, marks the maximum extension of its in situ stellar mass buildup.

We have shown in this work that the observed galaxy mass--size relation as defined by \citet{Trujillo2020} holds in cosmological hydrodynamic simulations of the formation of galaxies.
The remarkable agreement between simulations and observations can be used to infer the size and mass of dark matter haloes with an accuracy higher than 50\%.
The advantage of this definition of size is that the mass--size relation not only has reduced scatter but is also steeper for intermediate mass galaxies when compared to other empirical definitions of size, such as $R_{50}$.
This is what makes it a powerful (around six times better than $R_{50}$) predictor of galaxy and halo mass.

We stress that $R_1$ is only a proxy of the physically motivated definition of size, calibrated for late-type galaxies of mass similar to that of the the Milky Way.
On the other hand, as shown by \citet{Trujillo2020}, \citet{Arjona2025}, and this work, $R_1$ provides a consistently tight relation between size and stellar mass over many orders of magnitude in mass.
When applied to early-type galaxies, where the extent of in situ star formation has been modified by mergers, the threshold that defines the size should be between 3 and $10~\mathrm{M}_\odot\,\mathrm{pc}^{-2}$ \citep{Chamba2022}, suggesting that in situ star formation has happened at higher gas densities and larger redshift.

In a companion article, \citet{Arjona2025} have shown that simulations performed with different galaxy formation models all reproduce the same mass--size relation when $R_1$ defines the galaxy size. 
How the interplay of gas accretion, star formation, and stellar feedback yields to the same result for a variety of numerical models is not trivial to understand.
What all models have in common is that they are trying to reproduce the Kennicutt-Schmidt law and its threshold for star formation. 
Moreover, it seems that independent of the overall efficiency of star formation, galaxies of the same mass end up having the same size, although living in different haloes (see their figure~8, top panel).

The finding of \citet{Arjona2025} highlights a caveat for the prediction of the halo size. If the numerical model was not tuned to abundance matching models \citep[e.g.][]{Moster2013,Girelli2020} and galaxies were under- or over-massive with respect to the halo in which they were formed, the prediction of the halo size would suffer systematic errors.
It is, however, worth mentioning that empirical abundance matching models do not agree among them for intermediate- and low-mass galaxies, and the systematic difference would also depend on which model is adopted.
On the other hand, between the models, the difference in halo mass for a fixed stellar mass is less than a factor of two and slightly larger for different numerical models.

The question that remains open is why (simulated and observed) galaxies with stellar masses spanning many orders of magnitude and different morphologies follow such a tight mass--size relation.
\citet{SanchezAlmeida2020} investigated how the scatter in the mass--size relation significantly decreases when the size is measured at a fixed surface density, showing that galaxies with the same stellar mass always share at least one radius with identical surface density. \citet{Miller2019} hinted to this when defining the size of galaxies as the radius containing a given fraction of the stellar light, and they showed that when increasing the fraction of light, star forming and quiescent galaxies tend to move closer in the mass--size plane (see their figure~2).
None of these explanations are physical, and we will investigate this question further in the future.

\section{Conclusions}
\label{sec:conclusions}

With this work we have explored the relation between galaxy and halo size by applying the physically motivated definition of galaxy size of \citet{Trujillo2020} to simulated galaxies. We leveraged the public data of the EAGLE simulation project, namely, the galaxies produced in two simulated cosmological volumes of different resolutions, but calibrated to give similar global relations at $z=0$. In the following, we provide a summary of our conclusions:

\begin{itemize}
    \item We prove that the observed relation between galaxy stellar mass and size, as defined by \citet{Trujillo2020}, holds in cosmological, hydrodynamical simulations of galaxy formation over five orders of magnitude in stellar mass. The remarkable agreement between the simulations and observations can be used to precisely infer the galaxy stellar mass from its observed physical size \citep{Mowla2019,SanchezAlmeida2020}. We provide a fit to the distribution for galaxies more massive than $10^{8.5}\Mo$ (figure~\ref{fig:msra} and table~\ref{app:tab1}). We recovered the same scatter found by \citet{Trujillo2020}.
\vskip 6pt
    \item We propose the theoretical prediction that the size ($R_{200}$) and mass ($M_{200}$) of the dark matter halo hosting the galaxy can be accurately inferred by knowing the size of its central galaxy. Figure~\ref{fig:r1r200} and Table~\ref{app:tab2} quantify this prediction. The small scatter in the galaxy mass--size relation is a factor of two smaller than the observed scatter in the galaxy-to-halo size relation, whereas the slope of the relation greatly decrease the range of uncertainty when inferring the stellar mass of intermediate mass galaxies.
    These confirm the superiority of the galaxy size definition by nearly a factor of six when compared to observations using the half-mass radius \citep{Kravtsov2013}.
\end{itemize}

The origin of the relation and its connection to gas accretion, star formation, and stellar feedback is still poorly understood. More theoretical work will be dedicated to it in the near future.

\begin{acknowledgements}
This work has been supported by the Spanish Ministry of Science, Innovation and Universities (\textit{Ministerio de Ciencia, Innovación y Universidades}, MICIU) through the research grants PID2021-122603NB-C22 and PID2022-140869NB-I00. CDV also acknowledges support from MICIU in the early stages of development of this work through grants RYC-2015-18078 and PGC2018-094975-C22. This research made use of computing time on the high-performance computing systems \textsc{deimos} and \textsc{diva} of the \textit{Intituto de Astrof\'isica de Canarias}.
IT acknowledges support from: the ACIISI, Consejer\'{i}a de Econom\'{i}a, Conocimiento y Empleo del Gobierno de Canarias and the European Regional Development Fund (ERDF) under grant PROID2021010044; the IAC project P/302302, financed by the Ministry of Science and Innovation, through the State Budget, and by the Canary Islands Department of Economy, Knowledge, and Employment, through the Regional Budget of the Autonomous Community. This research also acknowledges support from the European Union through the following grants: UNDARK and `Excellence in Galaxies-Twinning the IAC' of the EU Horizon Europe Widening Actions programmes (project numbers 101159929 and 101158446). Funding for this work/research was provided by the European Union MSCA EDUCADO (GA 101119830). Views and opinions expressed are however those of the author(s) only and do not necessarily reflect those of the European Union or European Research Executive Agency (REA). Neither the European Union nor the granting authority can be held responsible for them.
The following Python packages have been used for this research:
\textsc{h5py}\footnote{\url{https://h5py.org}},
\textsc{numpy}\footnote{\url{https://numpy.org}} \citep{Harris2020},
\textsc{scipy}\footnote{\url{https://scipy.org}} \citep{Virtanen2020},
\textsc{matplotlib}\footnote{\url{https://matplotlib.org/}} \citep{Hunter2007}.
\end{acknowledgements}

\bibliographystyle{aa}
\bibliography{bibliography}

\appendix

\onecolumn

\section{Selection of the sample}
\label{app:sample}

\begin{figure*}
    \centering
    \includegraphics[width=0.32\textwidth]{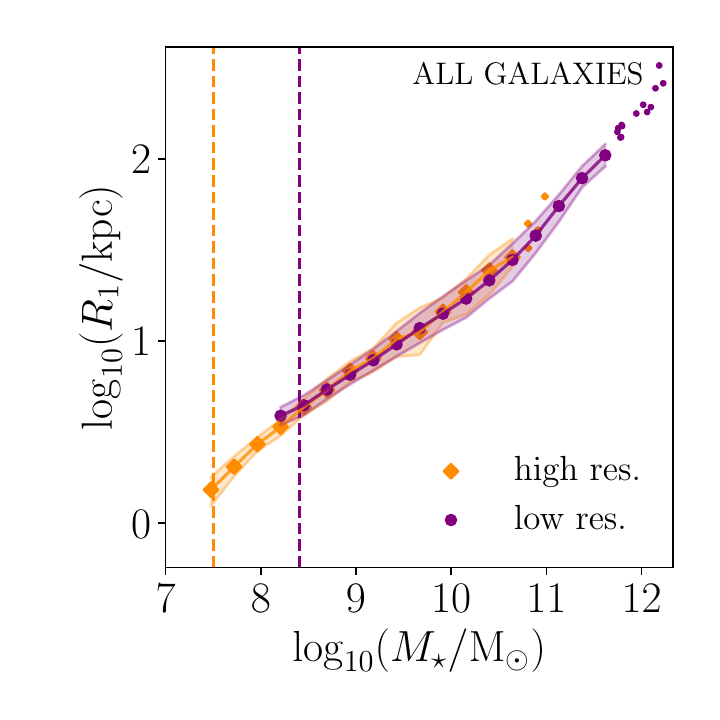}
    \includegraphics[width=0.32\textwidth]{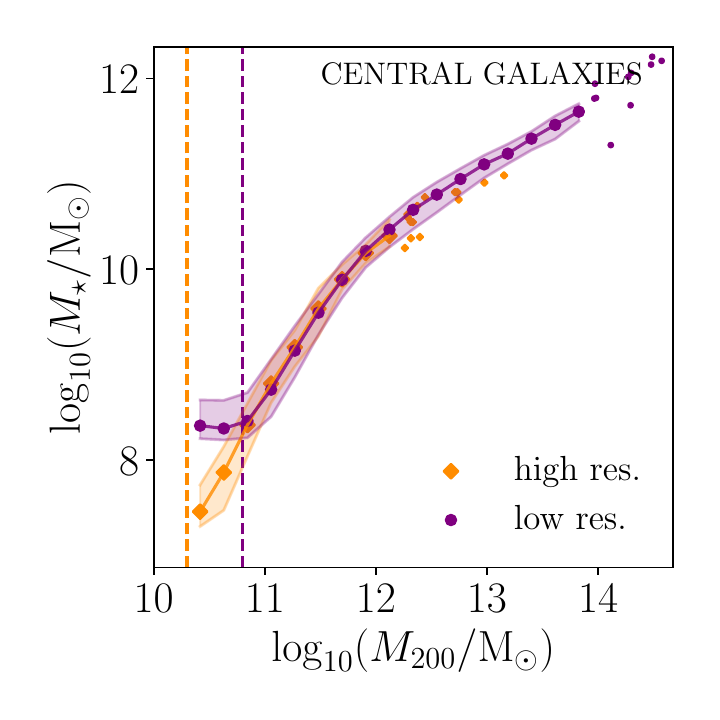}
    \includegraphics[width=0.32\textwidth]{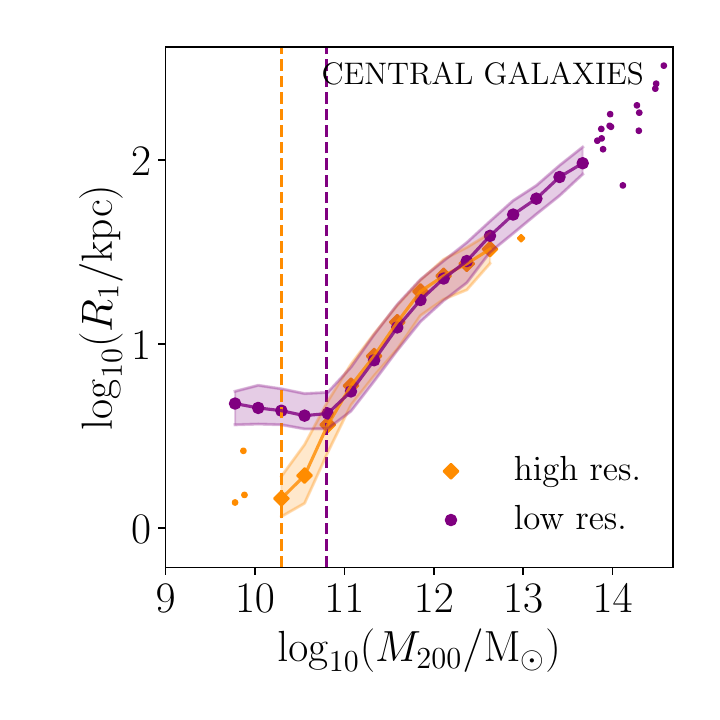}
    \caption{\textit{Left panel.} Relation between size and stellar mass for all galaxies in the two simulations. Dots and diamonds are the median values in bins of halo mass, and the shaded areas indicate the 16\% and 84\% quantiles. Single dots are for individual galaxies, in bins containing fewer than ten galaxies. The vertical lines mark the minimum masses considered for the two simulations in the computation of the analytic fits.
    \textit{Middle panel.} Relation between stellar mass, $M_\star$, and halo mass, $M_{200}$, for all central galaxies in the two simulations. Symbols are as in the left panel. The vertical lines mark the lower halo mass limit of the samples extracted from the two simulations and employed in the computation of the $R_1$--$R_{200}$ relation.
    \textit{Right panel.} Same as the middle panel, but for the $R_1$--$M_{200}$ relation.}
    \label{fig:resolution}
\end{figure*}

We show in figure~\ref{fig:resolution} the distribution of processed galaxies in the $(M_\star,R_1)$ (left panel, all galaxies), $(M_{200},M_\star)$ (mid panel, only central galaxies) and $(M_{200},R_1)$ (right panel, only central galaxies) planes.
The median values in bins of halo mass are shown as dots and diamonds for the low- and high-resolution simulations, respectively.
The shaded areas indicate the 16\% and 84\% quantiles.
Smaller dots and diamonds are for individual galaxies that are in bins with fewer than ten galaxies.
The two simulations overlap remarkably well over more than two orders of magnitude in both stellar mass and halo mass.
For the computation of the $R_1$--$M_\star$ relation, we selected galaxies with $M_\star>10^{7.5}\Mo$ from the recalibrated simulations, and with $M_\star>8\times10^{7.5}\Mo$ from the reference simulation, accordingly to the different mass resolution. These mass thresholds are depicted with vertical dashed lines in the left panel.
At the low-mass ends, the relations with $M_{200}$ (mid and right panels) flatten when reaching the resolution limit.
For the computation of the $R_1$--$R_{200}$ relation, we then consider central galaxies in haloes with mass above $M_\mathrm{thr}=10^{10.3}\Mo$ and $8\times M_\mathrm{thr}$ for the recalibrated and reference simulations, respectively, and accordingly to their particle mass resolution.
This choice is adequate for both $M_\star$--$M_{200}$ and $R_1$--$M_{200}$ relations.

\newpage

\section{Mass--size relation for central and satellite galaxies}
\label{app:censat}

\begin{figure*}
    \centering
    \includegraphics[width=0.45\textwidth]{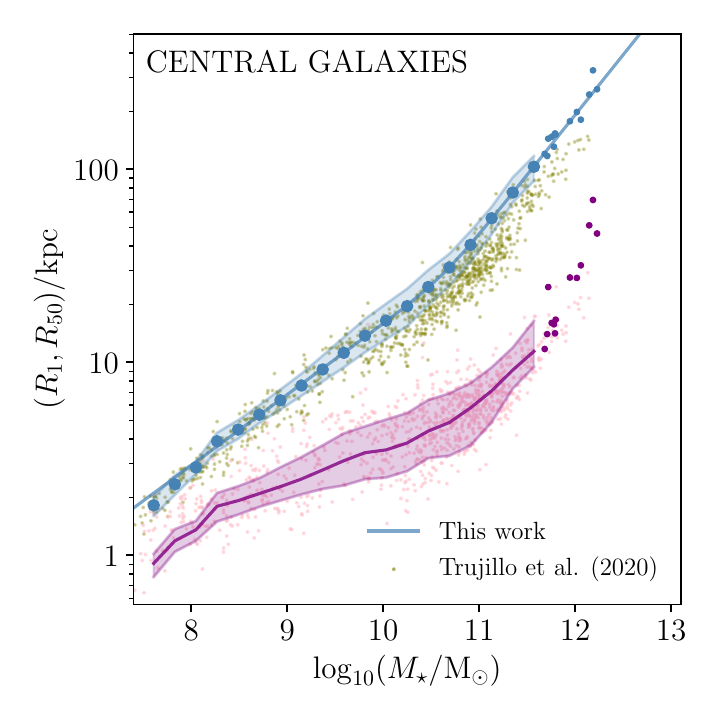}
    \includegraphics[width=0.45\textwidth]{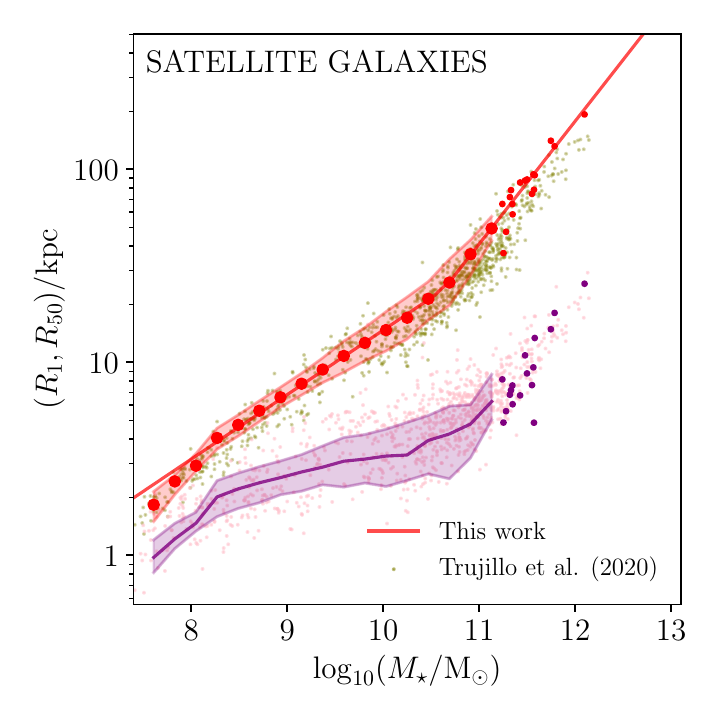}
    \caption{Correlation between $R_1$ and stellar mass for central (left panel) and satellite (right panel) galaxies in the sample. The median values for each bin are plotted together with the 16\% and 84\% quantiles (shaded area). For bins with fewer than ten galaxies, single values are plotted. Both fits to the data are for stellar masses $M_\star>10^{8.5}\Mo$. For comparison, we plot the relation for $R_{50}$ in purple.}
    \label{fig:msrc}
\end{figure*}

We restricted the sample to central and satellite galaxies, and repeated the analysis of section~\ref{sec:results}. We show in figure~\ref{fig:msrc} the stellar mass--size relations for central (left panel) and satellite (right panel) galaxies. The figure is similar to figure~\ref{fig:msra}, where median values are fitted with a double power-law (equation~\ref{eqn:powerlaw}) for stellar masses $M_\star>10^{8.5}\Mo$. For the same galaxies we also show the relation between stellar mass and half-mass radius, $R_{50}$, in purple. The median is given by the purple solid lines, whereas the shaded areas represent the 16\% and 84\% quantiles.
We report in table~\ref{app:tab1} the best-fit parameters for the stellar mass--size relation for all galaxies and for the subsamples of central and satellite galaxies, and in table~\ref{app:tab2} the best-fit parameters for the galaxy size-halo size relation for central galaxies. The errors on the parameters are given by the $\chi$-square minimisation. The dispersion around the best-fitting function is $\sigma_{\Delta\log_{10}R_1}=0.05$ and $\sigma_{\Delta\log_{10}R_1}=0.07$ for central and satellite galaxies, respectively.

\begin{table*}
\begin{center}
\begin{tabular}{|r|c|c|c|c|c|}
\multicolumn{6}{c}{GALAXY MASS--SIZE RELATIONS} \\
\hline
& $A/\mathrm{kpc}$ & $M_\mathrm{s}/\mathrm{M}_\odot$ & $\alpha$ & $\beta$ & $s$ \\
\hline
\hline
all galaxies & $22.8\pm 1.0$ & $(3.21\pm 0.35) \times 10^{10}$ & $0.35\pm 0.05$ & $0.60\pm 0.01$ & $4.25\pm 2.50$ \\
central galaxies & $27.1\pm 1.3$ & $(4.32\pm 0.57) \times 10^{10}$ & $0.37\pm 0.04$ & $0.62\pm 0.02$ & $2.48\pm 0.75$ \\
satellite galaxies & $25.4\pm 5.7$ & $(6.35\pm 0.48) \times 10^{10}$ & $0.31\pm 0.06$ & $0.75\pm 0.24$ & $1.70\pm 0.82$ \\
\hline
\end{tabular}
\end{center}
\caption{Best-fit parameters for the stellar mass--size relation for all galaxies and central and satellite galaxies. The errors are given by the $\chi$-square minimisation.}
\label{app:tab1}
\end{table*}

\begin{table*}
\begin{center}
\begin{tabular}{|r|c|c|c|c|c|}
\multicolumn{6}{c}{GALAXY SIZE--HALO SIZE RELATION} \\
\hline
& $A/\mathrm{kpc}$ & $R_\mathrm{s}/\mathrm{kpc}$ & $\alpha$ & $\beta$ & $s$ \\
\hline
\hline
central galaxies & $16.4\pm 1.4$ & $175.8\pm 10.4$ & $2.04\pm 0.07$ & $1.25\pm 0.03$ & $15.1\pm 17.9$ \\
\hline
\end{tabular}
\end{center}
\caption{Best-fit parameters for the galaxy size-halo size relation for central galaxies. The errors are given by the $\chi$-square minimisation.}
\label{app:tab2}
\end{table*}

\end{document}